\documentclass[11pt,reqno]{article}
\usepackage{graphics,graphicx,amsmath,amsthm,latexsym}
\begin{document}

\title{Perturbative quantum gauge invariance:\\
Where the ghosts come from.}

\author{Andreas Aste\\
Institute for Theoretical Physics\\
Klingelbergstrasse 82, Basel, Switzerland}
\date{January 23, 2003}
\maketitle

\begin{abstract}
A condensed introduction to quantum gauge theories is given
in the perturbative S-matrix framework; path integral
methods are used nowhere.
This approach emphasizes the fact that it is not necessary to start
from classical gauge theories which are then subject to quantization,
but it is also possible to recover the classical group structure
and coupling properties from purely quantum mechanical principles.
As a main tool we use a free field version of the
Becchi-Rouet-Stora-Tyutin gauge transformation, which
contains no interaction terms related to a coupling constant.
This free gauge transformation can be formulated in
an analogous way for quantum electrodynamics,
Yang-Mills theories with massless or massive gauge bosons
and quantum gravity.
\end{abstract}

\newpage

\section{Introduction}
It is well known that gauge theories play a fundamental role
in modern quantum field theory. From this observation one might
conclude that gauge invariance is a physical mechanism inherent
in many particle interactions. But this is not the case.
Gauge invariance is rather a mathematical artefact which stems from
the way we formulate (quantum) field theory, although its presence
is very helpful in order to achieve a consistent formulation
of many quantum field theories.
Like in the case of general relativity, where we are free to
choose the coordinate system according to our taste,
we also have some freedom how to deal with field theories.
The free quantum fields which are used in perturbation theory
in order to calculate physical quantities like e.g.~cross
sections simply provide a 'coordinatization' of
the problem under consideration, and they are 'as physical'
as a coordinate system.

One may illustrate this by a simple observation.
Massless particles with spin have only two degrees
of freedom, e.g. photons appear only in two helicity states.
Such states are called 'physical' in the
paper, whereas timelike or longitudinal photons, which
are an unavoidable byproduct of covariant quantization,
are called 'unphysical'. But if we take into account that one
can introduce running coupling constants and running masses
in interacting field theories, it becomes clear
that also 'physical' particles like quarks are merely
mathematical constructions, since we are always free to change
the structure of the Fock space underlying our
calculations by renormalizing the mass of the
quark asymptotic states. Therefore it is meaningless to ask
how a composite particle can be decomposed into free
'naked' particle states with a well-defined mass.
What one really should do is to calculate
vacuum expectation values of products of local
interacting fields in a nonperturbative
manner. From these distributions it is in principle possible
to reconstruct the true physical Hilbert space, as it is claimed
by the reconstruction theorem of Wightman \cite{Wightman}.
But this is a very difficult task and has been performed
only for exotic cases like e.g. quantum field theories in
1+1 spacetime dimensions \cite{Schwingermass,Abdalla,Walther,Adam}.

It is not the intention of this introduction to give a complete
description of all aspects related to the arbitraryness of
the formulation of quantum field theory, and for mathematical
details we must refer to the literature. But since we are
working in a strictly perturbative sense by using free field
operators only, the whole formalism used in this paper
can be formulated in a fully mathematical manner if necessary.

The structure of this paper is as follows:
First, we start with the free classical
electromagnetic field (or, more precisely, with
the classical gauge theory of a massless spin 1 field),
for which the basic issues of classical gauge invariance are explained.
Then the electromagnetic field is subject to quantization,
and we lift the gauge transformation on the quantum level.
It is explained how the quantum gauge transformation,
which is the free field analogue of the full
Becchi-Rouet-Stora-Tyutin (BRST) transformation
\cite{Becchi,Tyutin},
can be applied to quantum chromodynamics (QCD) along the same lines
as for the quantum electrodynamics (QED) case. As a further step, the method
is extended to massive gauge theories, where the Higgs field
is involved.
Finally, we outline how quantum gauge invariance can be formulated
for quantum gravity.

\section{The classical electromagnetic field}
The equation of motion for the noninteracting
vector potential $A^\mu$ in electrodynamics can be derived from
the Lagrangian
\begin{equation}
{\cal{L}}=-\frac{1}{4} F_{\mu \nu} F^{\mu \nu} = \frac{1}{2}
(\vec{E}^2-\vec{B}^2), \label{freeLag}
\end{equation}
where $F_{\mu \nu}=\partial_\mu A_\nu-\partial_\nu A_\mu$ is the
electromagnetic field strength tensor which can be expressed directly
in terms of the components of the electric and
magnetic fields $\vec{E}$ and $\vec{B}$. From the
Euler-Lagrange equations
\begin{equation}
\partial_\nu \frac{ \delta {\cal{L}}}{\delta \partial_\nu
A_\mu}-\frac{\delta {\cal{L}}}{\delta A_\mu}=0
\end{equation}
one derives then the wave equation
\begin{equation}
\partial_\nu \partial^\nu A_\mu-\partial_\mu \partial_\nu
A^\nu=0. \label{wave}
\end{equation}
A problem arises from the fact that the physically measurable
electromagnetic fields $\vec{E}$ and $\vec{B}$
remain unchanged if the vector potential is subject to a
gauge transformation expressed by a real scalar field $u$
($x=(x^0,\vec{x})$)
\begin{equation}
A_\mu(x) \rightarrow A_\mu(x)+\partial_\mu u(x), \label{gaugetrafo}
\end{equation}
i.e.~there is a redundance in the local description of the
electromagnetic fields by a vector potential, aside from global
topological information which may be contained in the potential
like in the case described by Aharonov and Bohm \cite{Aharonov}.
From the mathematical point of view, we generally assume that the
objects we are dealing with 'behave well', i.e.~$A_\mu$ and $u$
should be differentiable and behave at large spatial distances
in such a way that physical quantities like
the total energy of the field configuration are
finite.

The possibility to transform the vector potential without
changing its physical content allows to impose {\em{gauge
conditions}}. Due to its manifestly covariant form,
a very popular choice is the Lorentz gauge
condition, $\partial_\mu A^\mu=0$. This condition can be enforced by
transforming the initial vector potential according to (\ref{gaugetrafo})
with a field $u$ which fulfills
\begin{equation}
\Box u = \partial_\mu \partial^\mu u=-\partial_\mu A^\mu.
\end{equation}
Then the new transformed field satisfies the simple wave equation
\begin{equation}
\Box A^\mu=\partial_\nu \partial^\nu A^\mu=0. \label{simplewave}
\end{equation}
But the vector potential is not completely determined by the Lorentz
gauge condition. It is still possible to apply a gauge transformation
to the vector potential if $u$ fulfills the wave equation
$\partial_\nu \partial^\nu u=0$, such that the transformed vector
potential is still in Lorentz gauge. 
In the sequel we will denote the field $u$,
which is somehow related to the unphysical degrees of freedom of
the vector potential in Lorentz gauge, as 'ghost field'.
The reason will become clear at a later stage.

It is important to note that the equation of motion for the vector
potential can be manipulated by a modification of the Lagrangian
according to an early observation by Fermi \cite{Fermi}. Adding a
{\em{gauge fixing term}} to the Lagrangian
\begin{equation}
{\cal{L}}=
-\frac{1}{4} F_{\mu \nu} F^{\mu \nu} -\frac{1}{2} \xi
(\partial_\mu A^\mu)^2,
\end{equation}
where $\xi$ is an arbitrary real parameter, leads to the equation
of motion
\begin{equation}
\Box A_\mu-(1-\xi)\partial_\mu \partial_\nu
A^\nu=0.
\end{equation}
One may argue that the gauge fixing does not change the physical
content of the Lagrangian, since we may enforce the Lorentz gauge
condition $\partial_\mu A^\mu=0$, and then the gauge fixing term
vanishes in the Lagrangian.
For the special choice $\xi=1$ we recover the wave
equation (\ref{simplewave}). This choice is often referred to as
{\em{Feynman gauge}}, although this terminology is a bit misleading.
We must clearly distinguish between the gauge fixing,
which fixes the Lagrangian and determines the form of the wave
equation (e.g.~the Feynman gauge), and the choice of a gauge
condition for the vector potential (e.g.~the Lorentz gauge),
which reduces the redundance originating from the use of a
vector potential as a basic field. It is possible to impose
more general (nonlinear) gauge fixing conditions, e.g.~by adding a term
$\sim (\partial_\mu A^\mu)^4$ to the Lagrangian, but we restrict
ourselves to the most important cases here. We conclude this discussion
of the gauge fixing with the remark that it is indeed necessary
to add appropriate gauge fixing terms to the free Lagrangian
(\ref{freeLag}), since otherwise problems arise in the construction of
the free photon propagator.

Up to now, the discussion of the electromagnetic field has been
completely classical. In the next section, we will consider a
quantized version of the free electromagnetic field. For the field
operator $A_\mu(x)$
we will use the Feynman gauge such that the operator obeys
the simplest equation of motion (\ref{simplewave}).
All calculations could be performed
in a strictly analogous way for arbitrary $\xi-$gauges, but
the choice $\xi=1$ has obviously notational advantages \cite{Smatinv}.
Strictly speaking, each choice of $\xi$ leads after quantization
to a different quantum field theory with different Fock spaces and
field operators. But physical observables like cross sections
etc.~should be independent of the gauge fixing.

\section{Quantization of free fields}
It is well known that a free massless hermitian scalar field $\varphi(x)$
can be represented by
\begin{equation}
\varphi(t,\vec{x})=(2 \pi)^{-3/2} \int \frac{d^3k}{\sqrt{2 \omega}}
\Bigl(a(\vec{k}) e^{-i(\omega t -\vec{k} \vec{x})}
+a(\vec{k})^\dagger e^{i(\omega t -\vec{k} \vec{x})}  \Bigr),
\label{freefield}
\end{equation}
where $\omega=|\vec{k}|$ and the creation and annihilation operators
fulfill the commutation relations
\begin{equation}
[a(\vec{k}),a(\vec{k}')^\dagger]=\delta^{(3)}(\vec{k}-\vec{k}'),
\end{equation}
\begin{equation}
[a(\vec{k}),a(\vec{k}')]=[a(\vec{k})^\dagger,
a(\vec{k}')^\dagger]=0.
\end{equation}
The $^\dagger$ denotes the adjoint with respect to a positive
definite scalar product so that the operators can be represented in the
usual way in a Fock space with unique vacuum,
$\delta^{(3)}$ is the three dimensional Dirac distribution.

We now try to quantize the vector potential $A^\mu(x)$
as four independent real scalar fields
\begin{equation}
A^\mu(t,\vec{x})=(2 \pi)^{-3/2} \int \frac{d^3k}{\sqrt{2 \omega}}
\Bigl(a^\mu(\vec{k}) e^{-i(\omega t -\vec{k} \vec{x})}
+a^\mu(\vec{k})^\dagger e^{i(\omega t -\vec{k} \vec{x})}  \Bigr),
\label{naive}
\end{equation}
and naively we assume that the commutation relations for
the creation and annihilation operators are
\begin{displaymath}
[a^\mu(\vec{k}),a^\nu(\vec{k}')^\dagger]=\delta_{\mu \nu}
\delta^{(3)}(\vec{k}-\vec{k}'),
\end{displaymath}
\begin{equation}
[a^\mu(\vec{k}),a^\nu(\vec{k}')]=[a^\mu(\vec{k})^\dagger,
a^\nu(\vec{k}')^\dagger]=0. \label{creation}
\end{equation}
$\Box A^\mu=0$ is automatically satisfied
according to (\ref{naive}), since the spacetime dependence
of the field operator is given by the plane wave terms
$e^{\pm ikx}=e^{\pm i(\omega t -\vec{k} \vec{x})}$,
and $\delta$ is the Kronecker delta of the indices $\mu$, $\nu$.

A calculation shows that the free massless hermitian scalar field
$\varphi$ fulfills the commutation relation
\begin{equation}
[\varphi(x),\varphi(y)]=-i D(x-y),
\end{equation}
where $D(x-y)$ is the (massless) Pauli-Jordan distribution
\begin{equation}
D(x)=\frac{i}{(2 \pi)^3} \int
d^4 k \, \delta(k^2) \mbox{sgn}(k^0) e^{-ikx}
=\frac{1}{2 \pi} \mbox{sgn} (x^0) \delta(x^2).
\end{equation}
The Pauli-Jordan distribution fulfills the distributional
identity
\begin{equation}
\partial_0 D(x)|_{x_0=0}=\delta^{(3)}(\vec{x}) \label{timederiv}
\end{equation}
which implies the equal time commutation
relation for a scalar field $\varphi$ and its canonical
momentum $\pi=\dot{\varphi}$
\begin{equation}
[\varphi(x),\dot{\varphi}(y)]|_{x_0=y_0}=i \delta^{(3)}
(\vec{x}-\vec{y})
\end{equation}
which is often taken as the starting point for the quantization of
the scalar field. The distribution $D(x-y)$ was introduced by Pauli and
Jordan in 1928, when they performed the first covariant quantization
of the radiation field \cite{PauJo}.
The naive choice (\ref{naive}) comprises a problem, since the
commutation relations
\begin{equation}
[A^\mu(x),A^\nu(y)]=-i \delta_{\mu \nu} D(x-y)
\end{equation}
are not covariant. A simple way to remedy this defect is to
change the definition of $A^0$ into
\begin{equation}
A^0(x)=(2 \pi)^{-3/2} \int \frac{d^3k}{\sqrt{2 \omega}}
\Bigl( a^0 (\vec{k})e^{-ikx}-a^{0} (\vec{k})^\dagger e^{ikx}
\Bigr),
\end{equation}
i.e.~we make $A^0$ a skew-adjoint operator instead of self-adjoint.
Another strategy is to introduce a Fock space with negative norm
states, as proposed by Gupta and Bleuler \cite{Gupta,Bleuler}.
The commutation relations then become
\begin{equation}
[A^\mu(x),A^\nu(y)]=i g^{\mu \nu} D(x-y).
\end{equation}
We have introduced a nonhermitian field in order to save the
Lorentz symmetry of the theory, and therefore
one might expect problems with the unitarity of the
QED S-matrix. But no problems arise, a fact which is related
to the gauge symmetry of the theory.

A concluding remark is in order here. It is possible to avoid
unphysical polarization states by quantizing the photon field in
radiation gauge, which is expressed classically by the conditions
\begin{equation}
A^0=0, \quad \vec{\nabla} \cdot \vec{A}=0.
\end{equation}
The (quantized) photon field operator contains then
only physical (transverse) polarizations
\begin{equation}
A^0=0, \quad \vec{A}=(2 \pi)^{-3/2} \int \frac{d^3 k}{\sqrt{2 \omega}}
\sum_{\lambda=1,2} \vec{\epsilon} \, (\vec{k},\lambda)
[a(\vec{k},\lambda) e^{-ikx} + a^\dagger (\vec{k},\lambda)
e^{ikx}] \label{trans}
\end{equation}
with
\begin{equation}
\vec{\epsilon} \, (\vec{k},\lambda) \cdot \vec{\epsilon} \,
(\vec{k},\lambda') =\delta_{\lambda \lambda'} \quad  \mbox{and} \quad
\vec{\epsilon} \, (\vec{k},\lambda) \cdot \vec{k}=0
\end{equation}
and standard commutation relations for $a$ and $a^\dagger$.
But this is only seemingly an advantage because manifest Lorentz
symmetry is lost in (\ref{trans}), and no general strategy is
known how to deal with the renormalization of divergent higher order
contributions in interacting theories
without the guiding help of Lorentz symmetry.
Thus we have to make the choice between having unphysical
particles, but also manifest Lorentz covariance in our
calculations, or working only on a physical
Fock space, but without having a consistent scheme
at hand to regularize loop diagrams in perturbative calculations.

\section{A simple version of a quantum gauge transformation}
We introduce now a quantized version of the classical gauge transformation
for the free vector potential
\begin{equation}
A^\mu \rightarrow A^\mu+\partial^\mu u, \quad \Box u =0.
\end{equation}
Of course there exist different approaches for the treatment
of gauge invariance in quantum field theory, but the approach
presented here is quite simple and does the job.

As a first step, we define the gauge transformation operator
or {\em{gauge charge}} $Q$
\begin{equation}
Q=\int \limits_{x_0=const.} d^3 x \, \partial_\mu A^\mu(x)
\partial^{\! \! \! \! ^{^\leftrightarrow}}_0 u(x) . \label{charge}
\end{equation}
It is sufficient for the moment to consider $u$ as a real C-number field.
In the case of QCD, $u$ will necessarily become a fermionic scalar field.
It can be shown that Q is a well-defined operator on the Fock
space generated by the creation and annihilation operators
according to (\ref{creation}).
It is not important over which spacelike plane the integral
in (\ref{charge}) is taken, since $Q$ is time independent:
\begin{displaymath}
\dot{Q}=\int \limits_{x_0=const.} d^3 x \, 
(-\partial_0^2 \partial_\mu A^\mu u + \partial_\mu A^\mu
\partial_0^2 u)
\end{displaymath}
\begin{equation}
=\int \limits_{x_0=const.} d^3 x \,
(-\bigtriangleup \partial_\mu A^\mu u + \partial_\mu A^\mu
\bigtriangleup u)=0.
\end{equation}
This formal proof uses the wave equation and partial
integration.
Another way to show the time independence of the gauge
charge is to define the {\em{gauge current}}
\begin{equation}
j_g^\mu=\partial_\nu A^\nu \stackrel
{\leftrightarrow}{\partial^\mu} u, \quad Q=\int d^3x \, j^0_g,
\end{equation}
which is conserved
\begin{equation}
\partial_\mu j_g^\mu=\partial_\mu (\partial_\nu A^\nu \partial^\mu
u-\partial^\mu \partial_\nu A^\nu u)=0.
\end{equation}
A crucial property of the gauge charge is expressed by the
commutator with $A^\mu$, which is a C-number
\begin{equation}
[Q,A^\mu(x)]=i \partial^\mu u(x), \label{gaugecommutation}
\end{equation}
and all higher commutators like
\begin{equation}
[Q,[Q,A^\mu(x)]]=0.
\end{equation}
vanish for obvious reasons.
The reader is invited to check the commutation relation
(\ref{gaugecommutation}), an outline of the calculation can be found
in Appendix A.
Therefore we have
\begin{displaymath}
e^{-i \lambda Q} A^\mu e^{+i \lambda Q}=
A^\mu-\frac{i\lambda}{1 !} [Q,A^\mu]- \frac{\lambda^2}{2 !}
[Q,[Q,A^\mu]]+...
\end{displaymath}
\begin{equation}
=A^\mu-i \lambda [Q,A^\mu]=A^\mu+\lambda \partial^\mu u,
\label{gaugeseries}
\end{equation}
i.e.~$Q$ is a {\em{generator}} of gauge transformations.

\section{Definition of perturbative quantum gauge invariance}
We take the next step towards full QED and couple photons to
electrons.
In perturbative QED, the S-matrix is expanded as a power series
in the coupling constant $e$. At first order, the interaction
is described by the normally ordered product of free fields
\begin{equation}
{\cal{H}}_{\mbox{int}}(x) = -e:\bar{\Psi}(x)
\gamma^\mu \Psi(x): A_\mu (x),
\end{equation}
where $\Psi$ is the electron field operator.
The S-matrix is then usually given in the literature
by the formal expression ($T$
denotes time ordering)
\begin{displaymath}
S={\bf{1}}+\sum \limits_{n=1}^{\infty} \frac{(-i)^n}{n !}
\int d^4 x_1 \ldots d^4 x_n \, T[{\cal{H}}_{\mbox{int}}(x_1) \ldots
{\cal{H}}_{\mbox{int}}(x_n)]
\end{displaymath}
\begin{equation}
={\bf{1}}+\sum \limits_{n=1}^{\infty} \frac{1}{n !}
\int d^4 x_1 \ldots d^4 x_n \, T_n(x_1, \ldots x_n), \label{smatrix}
\end{equation}
where we have introduced the time-ordered products $T_n$
for notational simplicity, and we have
\begin{equation}
T_1(x)=-i {\cal{H}}_{\mbox{int}}(x)=ie:\bar{\Psi}(x)
\gamma^\mu \Psi(x): A_\mu (x).
\end{equation}
Expression (\ref{smatrix}) is plagued by
infrared and ultraviolet divergences (see Appendix E).
We leave this technical problem aside
and we assume that the $T_n$ are regularized, well-defined
operator valued distributions, which are symmetric in the
space coordinates $(x_1, \ldots x_n)$.

In the previous section, we used a real C-number field $u$
in order to show how the gauge transformation can be lifted on the
operator level.
We assume now that $u$ is a fermionic scalar field, and additionally
we introduce an anti-ghost field $\tilde{u}$, which is {\em{not}}
the hermitian conjugate of $u$.
The reason why
we choose $u$, $\tilde{u}$ to be fermionic, against all conventional
wisdom that
particles without spin are bosons, is very simple: Bosonic ghosts
do not allow to formulate a consistent theory in the case of
quantum chromodynamics, and therefore we start directly with
the 'correct' strategy. A second reason is that we want $Q$ to be
nilpotent; $Q^2=0$ which allows the definition of the physical
Hilbert space as
\begin{equation} 
{\cal{F}}_{phys}= ker \, Q/ran \, Q \label{physsubspace}
\end{equation}
(see Appendix C).

We give now a precise definition of
perturbative quantum gauge invariance for QED, which will work
in a completely analogous way for QCD.

Let
\begin {equation}
Q := \int d^3 x \, \partial_\mu A^\mu(x) \stackrel
{\leftrightarrow}{\partial_0} u(x) \label{Q}
\end{equation}
be the generator of (free field) gauge transformations,
called gauge charge for
brevity. This $Q$ has first been introduced in a famous paper by Kugo and 
Ojima \cite{11}.
The positive and negative frequency parts of the
free fields satisfy the $\{$anti-$\}$commutation relations
\begin{equation}
[A_\mu^{(\pm)}(x),A_\nu^{(\mp)}(y)]=ig_{\mu \nu} D^{(\mp)}(x-y),
\end{equation}
\begin{equation}
\{u^{(\pm)}(x),\tilde{u}^{(\mp)}(y)\}=-iD^{(\mp)}(x-y),
\end{equation}
and all other commutators vanish. Here, $D^{\mp}$ are the
positive and negative frequency parts of the massless
Pauli-Jordan distribution with Fourier transforms
\begin{equation}
{\hat{D}}^{(\pm)}(p)=\pm \frac{i}{2 \pi} \Theta(\pm p^0)
\delta(p^2).
\end{equation}
All the fields fulfill the
Klein-Gordon equation with zero mass, and as already mentioned, we are
working in Feynman gauge, but the following discussion would go through
with some technical changes in other covariant $\xi$-gauges as well.
Note that the quantum field $A^\mu$ does not satisfy
$\partial_\mu A^\mu=0$ on the whole Fock space, but only on the
physical subspace.
An explicit representation of the ghost fields is given by
\begin{equation}
u(x)=(2 \pi)^{-3/2} \int  \frac{d^3 k}{\sqrt{2 \omega}}
\Bigl(c_2(\vec k)e^{-ikx}+c_1^\dagger (\vec k)e^{ikx}\Bigl),
\end{equation}
\begin{equation}
\tilde u(x)=(2 \pi)^{-3/2} \int  \frac{d^3 k}{\sqrt{2 \omega}}
\Bigl(-c_1(\vec k)e^{-ikx}+c_2^\dagger
(\vec k)e^{ikx}\Bigl),
\end{equation}
\begin{equation}
\{c_i(\vec k),c_j^\dagger (\vec k')\}=
\delta_{ij}\delta^{(3)}(\vec k-\vec k'),
\quad i,j=1,2,
\end{equation}

In order to see how the
infinitesimal gauge transformation acts on the free fields, we have to
calculate the commutators (see \cite{5} and Appendix A)
\begin{equation}
[Q,A_\mu]=i\partial_\mu u  ,
\quad \{Q,u\}=0  , \quad \{Q,\tilde{u}\}=-i\partial_\nu A^\nu  ,
\quad [Q,\Psi]=[Q,{\bar{\Psi}}]=0 . \label{commu}
\end{equation}
The commutators of $Q$ with the electron field are of course trivial,
since the operators act on different Fock space sectors.
We need only the first and the last commutator in (\ref{commu}) here,
the others will become important in the QCD case.
From (\ref{gaugeseries}) we know that the commutator of $Q$ with an
operator gives the first order variation of the operator subject to a
gauge transformation.
Then we have for the first order interaction $T_1$
\begin{displaymath}
[Q,T_1(x)] = -e:\bar{\Psi} (x) \gamma^\mu \Psi (x) : \partial_\mu u(x)
\end{displaymath}
\begin{equation}
=i \partial_\mu (ie:\bar{\Psi} (x) \gamma^\mu \Psi (x) :u(x))
= i \partial_\mu T^\mu_{1/1} (x). \label{gi}
\end{equation}
Here we used the fact that the electron current is conserved
\begin{equation}
\partial_\mu :\bar{\Psi} \gamma^\mu \Psi:=0.
\end{equation}
because $\Psi$ fulfills the free Dirac equation.
Note that the free electron field is {\em{not}} affected by the
gauge transformation.
We may call $T^\mu_{1/1}=ie:\bar{\Psi} \gamma^\mu \Psi:u$
the 'Q-vertex' of QED.
The generalization of (\ref{gi}) to n-th order is
\begin{equation}
[Q,T_n(x_1,...x_n)] = i \sum_{l=1}^{n} \partial_\mu^{x_l} T_{n/l}^{\mu}
(x_1,...x_n) = (\hbox{\rm sum of divergences}) \quad , \label{dive}
\end{equation}
where $T^\mu_{n/l}$ is again a mathematically well-defined version of
the time-ordered
product
\begin{equation}
T^\mu_{n/l}(x_1,...,x_n)
\, "=" \, T(T_1(x_1)...T^\mu_{1/1} (x_l)...T_1(x_n)).
\end{equation}
It is relatively easy to see how (\ref{dive}) comes about
if one understands the second order case. This example is discussed in
detail in Appendix D.
We {\em{define}} (\ref{dive}) to be the condition of gauge invariance 
\cite{5}.

If we consider for a fixed $x_l$ all terms in $T_n$ 
with the external field operator $A_\mu(x_l)$
\begin{equation}
T_n(x_1,...x_n) = :t^\mu_l(x_1,...x_n) A_\mu(x_l):+...
\end{equation}
(the dots represent terms without $A_\mu(x_l)$),
then gauge invariance (\ref{dive}) requires
\begin{equation}
\partial_\mu^l [t^\mu_l(x_1,...x_n)u(x_l)]=t^\mu_l(x_1,...x_n) \partial_\mu
u(x_l)
\end{equation}
or
\begin{equation}
\partial_\mu^l t^\mu_l(x_1,...x_n) = 0, {\label{eich}}
\end{equation}
i.e.~we obtain the Ward-Takahashi identities \cite{13} for QED.
The Ward-Takahashi identities express the implications of
gauge invariance of QED, which is defined here on the
operator level, by C-number identities for Green's distributions.

We have found the following important property of QED: There exists
a symmetry transformation generated by the gauge charge $Q$, which
leaves the S-matrix elements invariant, since the gauge transformation
only adds divergences in the analytic sense
to the S-matrix expansion which vanish after integration
over the coordinates $x_1, \dots x_n$ (see Appendix E).

The S-matrix 'lives' on a Fock space ${\cal{F}}$ which contains physical
and unphysical states.
E.g., unphysical single photon states can be created
by acting on the perturbative vacuum $|\Omega \rangle$ with
timelike or longitudinal creation operators:
\begin{equation}
a^{0}(\vec{k})^\dagger |\Omega \rangle, \quad
\sum \limits_{j=1,2,3} {k_j}a^{j} (\vec{k})^\dagger |\Omega \rangle,
\end{equation}
and states which contain ghosts are always unphysical.
The physical subspace ${\cal{F}}_{phys}$ is the space of
states $|\Phi \rangle$
without timelike and longitudinal photons and ghost, that means
\begin{equation}
a^0(\vec{k})|\Phi \rangle=0, \quad {k_j}a^{j}(\vec{k})
|\Phi \rangle=0, \quad c_{1,2} (\vec{k})|\Phi \rangle=0 \quad \forall
\vec{k}. \label{physicalcond}
\end{equation}
Note that the definition of the physical space is
{\em{not}} Lorentz invariant, but the definition (\ref{physsubspace})
is.

The observation that QED is gauge invariant is interesting,
but the true importance of gauge invariance is the fact
that it allows to prove on a formal level
the {\em{unitarity}} of the S-matrix on the physical subspace
(see the last paper of \cite{5}).
Due to the presence of
the skew-adjoint operator $A^0$ or the presence of unphysical
longitudinal and timelike photon states, the S-matrix is
not unitary on the full Fock space, but it is on ${\cal{F}}_{phys}$.
We do not give the algebraic proof here, but we point out that
gauge invariance is the basic prerequisite which ensures
unitarity. A detailed discussion of this fact can be found in
\cite{5,Razumov,pgi}.
We have introduced the ghosts only as a formal tool, since
they 'blow up' the Fock space unnecessarily, and they do not
interact with the electrons and photons. But in QCD, the situation
is not so trivial.

\section{Quantum chromodynamics}
As a further step, we consider QCD without fermions
(i.e.~quarks) now.
The first order coupling of the gluons that we obtain from the
classical QCD Lagrangian is given by
(see Appendix B)
\begin{displaymath}
T_1^A(x)=i \frac{g}{2} f_{abc} : A_{\mu a}(x) A_{\nu b}(x)
F^{\nu \mu}_c (x):
\end{displaymath}
\begin{equation}
=igf_{abc} : A_{\mu a}(x) A_{\nu b}(x)
\partial^\nu A^\mu_c (x): , \quad a,b,c=1,\ldots 8
\label{gluonQCD}
\end{equation}
where $F^{\mu \nu}_c=\partial^\mu A^\nu_c-\partial^\nu A^\mu_c$
and $f_{abc}$ are the totally antisymmetric structure constants
of the gauge group SU(3).
The asymptotic free fields satisfy the commutation relations
\begin{equation}
[A_{\mu a}^{(\pm)}(x),A_{\nu b}^{(\mp)}(y)]=i \delta_{ab}
g_{\mu \nu} D^{(\mp)}(x-y)
\end{equation}
and
\begin{equation}
\{u_a^{(\pm)}(x),\tilde{u}_b^{(\mp)}(y)\}=-i \delta_{ab} D^{(\mp)}(x-y),
\end{equation}
and all other \{anti-\}commutators vanish. 
Defining the gauge charge as in (\ref{Q}) by
\begin{equation}
Q:=\int d^3x \, \partial_\mu A^\mu_a (x)  \stackrel
{\leftrightarrow}{\partial_0} u_a(x) ,
\end{equation}
where summation over repeated indices is understood,
we are led to the following commutators with the fields:
\begin{equation}
[Q,A_a^\mu]=i\partial^\mu u_a ,
\quad [Q,F_a^{\mu \nu}]=0 , \quad
\{Q,u_a\}=0 , \quad \{Q,\tilde{u}_a\}= -i\partial_\mu A^\mu_a \quad .
\label{comm}
\end{equation}
For the commutator of $Q$ with $T_1^A$ we obtain
\begin{displaymath}
[Q,T_1^A(x)]=ig f_{abc} :\{[Q,A_{\mu a}] A_{\nu b} \partial^\nu A^\mu_c+
A_{\mu a} [Q,A_{\nu b}] \partial^\nu A^\mu_c+ A_{\mu a} A_{\nu b}
[Q, \partial^\nu A^\mu_c]\}:
\end{displaymath}
\begin{equation}
=-gf_{abc} :\{ \partial_\mu u_a A_{\nu b} \partial^\nu A^\mu_c+
\partial_\nu (A_{\mu a} u_b \partial^\nu A^\mu_c)\}:. \label{nondiv}
\end{equation}
The last term is a divergence, but the first term spoils gauge
invariance and therefore the unitarity of the theory.
To restore gauge invariance, we must somehow compensate the
first term in (\ref{nondiv}).

We consider therefore the gluon-ghost coupling term
\begin{equation}
T_1^u(x)=igf_{abc}
:A_{\mu a}(x) u_b (x) \partial^{\mu} {\tilde{u}}_c(x): ,
\label{ghostQCD}
\end{equation}
For the commutator with $Q$ we get
\begin{displaymath}
[Q,T_1^u]=ig :[Q,A_{\mu a}] u_b \partial^{\mu} {\tilde{u}}_c
+A_{\mu a} \{ Q,u_b \} \partial^{\mu} {\tilde{u}}_c
-A_{\mu a} u_b \partial^{\mu} \{ Q, {\tilde{u}}_c \}:
\end{displaymath}
\begin{equation}
=-g f_{abc} :\{ \partial_\mu u_a u_b \partial^\mu \tilde{u}_c
+A_{\mu a} u_b \partial^\mu \partial^\nu A_{\nu c} \}:.
\end{equation}
Note that we can always decompose the commutator in a clever
way such that only anticommutators of $Q$ with the ghost fields
and commutators with the gauge fields appear.
Taking the antisymmetry of $f_{abc}$
and $:u_a u_b:=-:u_b u_a:$ into account, we see that the
first term is a divergence
\begin{equation}
f_{abc}:\partial_\mu u_a u_b \partial^\mu \tilde{u}_c:
=\frac{1}{2} f_{abc} \partial_\mu :u_a u_b \partial^\mu \tilde{u}_c:
\end{equation}
because $\tilde{u}_c(x)$ fulfills the wave equation.
The second term can be written as
\begin{equation}
-g f_{abc} :A_{\mu a} u_b \partial^\mu \partial^\nu A_{\nu c}:=
-g f_{abc} :\partial^\nu(A_{\mu a} u_b \partial^\mu A_{\nu c})-
\partial^\nu u_b A_{\mu a} \partial^\mu A_{\nu c}:.
\end{equation}
Interchanging $a \leftrightarrow b$ and $\mu \leftrightarrow \nu$
in the last term, it becomes equal to the first term in (\ref{nondiv}).

Hence, the combination
\begin{equation}
T_1(x)=igf_{abc} \{\frac{1}{2}: A_{\mu a}(x) A_{\nu b}(x) F^{\nu \mu}_c (x):
-:A_{\mu a}(x) u_b (x) \partial^{\mu} {\tilde{u}}_c(x): \}
\label{QCD}
\end{equation}
leads to a gauge invariant first order coupling
\begin{equation}
[Q,T_1]=g f_{abc} : -\partial_\nu (A_{\mu a} u_b
(\partial^\nu A^\mu_c-\partial^\mu A^\nu _c)+\frac{1}{2}
\partial_\nu (u_a u_b \partial^\nu \tilde{u}_c):.
\end{equation}

In contrast to QED, the nonlinear selfinteraction of the
unphysical degrees of freedom of the gluons spoils gauge invariance.
Coupling ghosts in an appropriate way to the gluons restores the
consistency of the theory. Thats why we need ghosts. They interfere
destructively with the unphysical gluons and save the unitarity
of the theory.

The first order coupling given in (\ref{QCD}) is in fact not the most
general one, and it is also possible to construct a gauge invariant
first order coupling with bosonic ghosts \cite{5}; but gauge invariance
then breaks down at second order of perturbation theory. Gauge
invariance at order $n$ is again defined by (\ref{dive}).
We have discussed gauge invariance only at first order here.
To prove that gauge invariance (and other symmetries of the
theory) is not broken by renormalization, is the 'hard problem'
of renormalization theory \cite{Dutsch1,Hurth1,Hurth2}.

The 4-gluon term $\sim g^2$ which also appears in
the classical Lagrangian is missing in $T_1$. This term appears
as a necessary local normalization term at second order, and its
structure is also fixed by perturbative gauge invariance.
This stresses the fact that perturbative gauge invariance
is strongly related to the formal expansion of the theory in powers of
the coupling constant g.

We started this section by presupposing that the coupling of
the gluons is already given by the classical Lagrangian.
But perturbative gauge invariance is indeed a very restrictive
condition: It fixes the interaction to a large extent.
We outline here how this fact can be derived. Only a part of
the derivation is given here, and it is not assumed that the
reader will check the calculations in detail. A full discussion
is given in \cite{pgi}, and
for further reading we recommend also \cite{Ghoststory}.

We start by describing the interaction of the massless gluons
by the most general renormalizable ansatz (i.e.~the dimension of the
interaction terms is energy$^4$)
with zero ghost number (coupling with non-zero
ghost number would affect the theory only on the unphysical sector)
\begin{eqnarray}
\tilde{T}_1(x) & = ig \Bigl\{ 
& \tilde{f}^1_{abc} : A_{\mu a}(x) A_{\nu b}(x)
\partial^\nu A^\mu_c (x): +
\nonumber \\
& & \tilde{f}^2_{abc} : A_{\mu a} u_b \partial^\mu \tilde{u}_c : +
\nonumber \\
& & \tilde{f}^3_{abc} : A_{\mu a} \partial^\mu u_b \tilde{u}_c : +
\nonumber \\
& & \tilde{f}^4_{abc} : A_{\mu a} A^\mu_b \partial_\nu A ^\nu_c: +
\nonumber \\
& & \tilde{f}^5_{abc} : \partial_\nu A^\nu_a u_b \tilde{u}_c : \Bigr\}
\quad , \quad  \tilde{f}^4_{abc} = \tilde{f}^4_{bac} ,\nonumber
\end{eqnarray}
where the $\tilde{f}$'s are arbitrary real constants. This
first order coupling term is then antisymmetric with respect
to the conjugation $K$ defined in Appendix C.

Adding divergence terms to $\tilde{T}_1$ will not change the
physics of the theory, since divergences get integrated out
in (\ref{smatrix}), and we assume that this property is not destroyed
by the renormalization of the theory. Furthermore, the (anti-)commutator
of $Q$ with a divergence is also a divergence. 
We can therefore always calculate modulo
divergences in the following. We
modify $\tilde{T}_1$ by adding $-\frac{ig}{4}
(\tilde{f}^1_{abc}+\tilde{f}^1_{cba}) \partial_\nu : A_{\mu a} A_{\nu b}
A^\mu_c :$ and $-ig \tilde{f}^3_{abc} \partial_\mu : A^\mu_a u_b \tilde{u}_c:$
and arrive at a more compact, equivalent first order coupling
\begin{eqnarray}
T_1 & = ig \bigl\{ 
& f^1_{abc} : A_{\mu a} A_{\nu b} \partial^\nu A^\mu_c: +
\nonumber \\
& & f^2_{abc} : A_{\mu a} u_b \partial^\mu \tilde{u}_c : +
\nonumber \\
& & f^4_{abc} : A_{\mu a} A^\mu_b \partial_\nu A^\nu_c : +
\nonumber \\
& & f^5_{abc} : \partial_\nu A^\nu_a u_b \tilde{u}_c : \bigr\} ,
\label{nr35}
\end{eqnarray}
where 
\begin{equation}
f^1_{abc}=-f^1_{cba} \quad f^4_{abc}=f^4_{bac} . \label{asy}
\end{equation}
The gauge variation of $T_1$ is
\begin{eqnarray}
[Q,T_1] & = -g f^1_{abc} & \Bigl\{
: \partial_\mu u_a A_{\nu b} \partial^\nu
A^\mu_c +
\nonumber \\
& & A_{\mu a} \partial_\nu u_b \partial^\nu A^\mu_c +
\nonumber \\
& & A_{\mu a} A_{\nu b} \partial^\nu \partial^\mu u_c : \Bigr\}+
\nonumber \\
& -g f^2_{abc} & \Bigl\{: \partial_\mu u_a u_b \partial^\mu \tilde{u}_c +
\nonumber \\
& & A_{\mu a} u_b \partial^\mu \partial_\nu A^\nu_c: \Bigr\} +
\nonumber \\
& -2g f^4_{abc} & :\partial_\mu u_a A^\mu_b \partial_\nu A^\nu_c : +
\nonumber \\
& -g f^5_{abc} & :\partial_\nu A^\nu_a u_b \partial_\mu A^\mu_c : .
\label{var}
\end{eqnarray}
Gauge invariance requires that the gauge variation be a divergence,
hence we make again a general ansatz
\begin{eqnarray}
[Q,T_1] & = g\partial_\mu \bigl\{ & g^1_{abc} :\partial^\mu u_a
A_{\nu b} A^\nu_c : +
\nonumber \\
& & g^2_{abc} : u_a A^\mu_b \partial_\nu A^\nu_c : +
\nonumber \\
& & g^3_{abc} : \partial^\mu u_a u_b \tilde{u}_c : +
\nonumber \\
& & g^4_{abc} : u_a u_b \partial^\mu \tilde{u}_c : +
\nonumber \\
& & g^5_{abc} : \partial_\nu u_a A^\nu_b A^\mu_c : +
\nonumber \\
& & g^6_{abc} : u_a \partial^\mu A^\nu_b A_{\nu c} : +
\nonumber \\
& & g^7_{abc} : u_a \partial^\nu A^\mu_b A_{\nu c} : 
\bigr\} , \label{div}
\end{eqnarray}
where $g^ 1_{abc}=g^ 1_{acb}, g^ 4_{abc}=-g^ 4_{bac}$.
Comparing the terms in (\ref{var}) and (\ref{div}), we obtain
a set of constraints for the coupling coefficients:
\begin{eqnarray}
-f^1_{cab} = 2 g^1_{abc} + g^6_{abc} & \mbox{from} & :\partial^\mu u
\partial_\mu A_\nu A^\nu :
\label{1} \\
g^2_{abc} + g^5_{abc} = -2f^4_{abc}
& & :\partial^\mu u A_\mu \partial_\nu A^\nu: \label{2} \\
g^2_{abc} + g^2_{acb} = -f^5_{bac}-f^5_{cab}
& & :u \partial_\nu A^\nu \partial_\mu A^\mu :
\label{3} \\
-f^2_{abc} = g^3_{abc} + 2 g^4_{abc} & & :\partial_\mu u u
\partial^\mu \tilde{u}:
\label{4} \\ 
g^3_{abc} = g^3_{bac} & & :\partial_\mu  u \partial^\mu u \tilde{u}: 
\label{5} \\
-f^1_{cba} - f^1_{bca} = g^5_{abc} + g^5_{acb} & & :\partial_\mu
\partial_\nu u A^\nu A^\mu:
\label{6} \\
-f^1_{acb} = g^5_{abc} + g^7_{abc} & &  :\partial_\mu u \partial_\nu
A^\mu A^\nu:
\label{7} \\
g^6_{abc}= - g^6_{acb} & & :u \partial_\mu A_\nu \partial^\mu A^\nu:
\label{8} \\
-f^2_{cab} = g^2_{acb} +g^7_{abc} & & :u \partial_\mu \partial_\nu
A^\mu A^\nu:
\label{9} \\
g^7_{abc}= - g^7_{acb} & &: u \partial_\nu A^\mu \partial^\mu A^\nu:
\label{10} .
\end{eqnarray}
From (\ref{6}), (\ref{7}), and (\ref{10}) we readily derive
\begin{equation}
f^1_{abc} + f^1_{acb} - f^1_{bca} - f^1_{cba} =0 .
\end{equation}
Combining this with (\ref{asy}), 
we obtain the first important result that $f^1_{abc}$ is
{\em{totally antisymmetric}}.
$g^1_{abc}$ is symmetric in $b$ and $c$; from (\ref{8}) and (\ref{1}) we
conclude $g^1_{abc}=0$ and $g^6_{abc}=-f^1_{abc}$.

We did not yet fully exploit gauge invariance.
Taking into account additionally that the gauge charge is
nilpotent $Q^2=0$ and consequently
\begin{equation}
\{Q,[Q,T_1]\}=0
\end{equation}
provides further relations which fix the interaction
to
\begin{equation}
T_1=T_1^{YM} + T_1^{D}  ,
\end{equation}
\begin{equation}
T_1^{YM} = ig f^1_{abc} \bigl\{
:\frac{1}{2} A_{\mu a} A_{\nu b} F^{\nu \mu}_c:
-:A_{\mu a} u_b \partial^\mu \tilde{u}_c: \bigr\} ,
\end{equation}
\begin{equation}
T_1^{D} = igf^{5}_{abc}
:\partial_\nu A^\nu_a u_b \tilde{u}_c: ,
\end{equation}
where 'YM' stands for 'Yang-Mills' and 'D' for 'Deformation', 
and $f^{5}_{abc}=-f^{5}_{cba}$.

We stop our analysis here for the sake of brevity.
A detailed analysis of gauge invariance {\em{at second order}}
shows \cite{pgi}
that the coupling coefficients $f_{abc}$ must fulfill the Jacobi identity
\begin{equation}
f_{abc} f_{dec}+f_{adc} f_{ebc} + f_{aec} f_{bdc} = 0,
\end{equation}
i.e.~we are lead to the nice result that the $f_{abc}$ are structure
constants of a Lie group. Note that the Lie structure of
gauge theories is usually presupposed, but here it follows
from basic symmetries of quantum field theory, namely
quantum gauge invariance (which is related to unitarity) and
Lorentz covariance.

Finally, we are left with the term
\begin{equation}
T_1^{D} = igf^{5}_{abc}
:\partial_\nu A^\nu_a u_b \tilde{u}_c:. \label{cobo}
\end{equation}
One can show that this interaction term which contains
only unphysical fields does not contribute to the S-matrix on
the physical sector.

The reason why gauge invariance at higher orders of the perturbation
expansion is a nontrivial task can be understood qualitatively
on a basic level.
Commutators of free fields can be expressed by (derivatives of)
Pauli-Jordan distributions, e.g. we have
\begin{displaymath}
\{ \partial_\nu A^\nu(x),\partial_\mu A^\mu(y)]=
i g^{\nu \mu} \partial_\nu^x \partial^y_\mu D(x-y)
\end{displaymath}
\begin{equation}
=-i \Box_x D(x-y)=0.
\end{equation}
But through the timeordering process, 
Feynman type propagators $\sim D_F(x-y)$
appear in the amplitudes instead of Pauli-Jordan distributions,
which fulfill the inhomogeneous wave equation
\begin{equation}
\Box D_F(z)=-\delta^{(4)}(z),
\end{equation}
i.e. derivatives acting on Feynman propagators may
generate 'anomalous' local terms, because time ordering does
not commute with analytic derivation.
Therefore, maintaining gauge invariance may imply restrictions to the
coupling structure also at higher orders. If it is not
possible to fix the theory such that gauge invariance
can be maintained, the theory is called {\em{nonrenormalizable}}.
The result obtained by the analysis above
can be considered as an inversion of 't Hooft's famous result
that Yang-Mills theories are renormalizable \cite{tHooft}.

In this section, we focused on the purely gluonic part of QCD.
A full discussion would include also the coupling of gluons to
quarks, but all important features are contained in
the present discussion.

\section{Yang-Mills theories with massive gauge fields}
In this section we give a short discussion of a general Yang-Mills
theory with massive gauge bosons. Free massive spin 1 fields with color
index $a$ satisfy the Proca equation
\begin{equation}
\Box A^\mu_a -\partial^\mu (\partial_\nu A^\nu_a) +
m_a^2 A^\mu_a =0. \label{proca}
\end{equation}
Taking the divergence of (\ref{proca}) shows that the field
automatically satisfies the Lorentz condition $\partial_\mu
A^\mu_a=0$.
Again we may quantize the fields as four independent scalar fields
which fulfill the wave equation
\begin{equation}
(\Box + m_a^2) A^\mu_a =0,
\end{equation}
and the Lorentz condition shall be enforced on the Fock space
like in the massless case by selecting a physical subspace
which contains the three allowed polarizations.
The definition of the gauge charge via (\ref{Q}) does no longer lead to
a nilpotent gauge charge due to the mass of the gauge field.
This can be restored by introducing scalar bosonic fields
$\Phi_a$ with the same mass as the corresponding gauge field
\begin{equation}
(\Box + m^2_a) \Phi_a =0 .
\end{equation}
These fields are called Higgs ghosts or would-be Goldstone bosons, and they
are unphysical.
We demand that the free fields then satisfy the commutation relations
\begin{equation}
[A^\mu_a(x),A^\nu_b(y)]=i \delta_{ab} g^{\mu \nu} D_{m_a} (x-y),
\end{equation}
\begin{equation}
[u_a(x),\tilde{u}_b (y)]=-i \delta_{ab} D_{m_a} (x-y),
\end{equation}
\begin{equation}
[\Phi_a(x),\Phi_b (y)]=-i \delta_{ab} D_{m_a} (x-y),
\end{equation}
where $D_{m_a} (x-y)$ are the usual Pauli-Jordan distributions
for massive fields
\begin{equation}
D_m(x)=\frac{i}{(2 \pi)^3} \int d^4 k \,
\delta(k^2-m^2) \mbox{sgn} (k^0) e^{-ikx} .
\end{equation}
Then we can define a nilpotent gauge charge by
\begin{equation}
Q := \int d^3 x \, (\partial_\mu A^\mu_a(x)
+m_a \Phi_a(x))\stackrel
{\leftrightarrow}{\partial_0} u(x) \label{Qmass}
\end{equation}
Gauge variations are then obtained from the commutators
\begin{displaymath}
[Q,A^\mu_a]=i \partial^\mu u_a, \quad \{Q,u_a \}=0,
\end{displaymath}
\begin{equation}
\{Q,\tilde{u}_a\}=
-i(\partial_\mu A^\mu_a +m_a \Phi_a), \quad [Q,\Phi_a]=i m_a u_a.
\label{gaugemass}
\end{equation}
The coupling (\ref{QCD}) which we obtained in the massless case
\begin{equation}
T_1(x)=igf_{abc} \{\frac{1}{2}: A_{\mu a}(x) A_{\nu b}(x) F^{\nu \mu}_c (x):
-:A_{\mu a}(x) u_b (x) \partial^{\mu} {\tilde{u}}_c(x): \}
\end{equation}
is not gauge invariant anymore due to the additional massterms in the
gauge variations (\ref{gaugemass}).
But gauge invariance can be restored at least at first order if we add
to $T_1$ a Higgs ghost coupling
\begin{displaymath}
T_1^\Phi=i \frac{g}{2} f_{abc} \Bigl(
\frac{m_b^2+m_c^2-m_a^2}{m_b m_c}: A_{\mu a} \Phi_b \partial^\mu
\Phi_c:
\end{displaymath}
\begin{equation}
+\frac{m_a^2+m_c^2-m_b^2}{m_a} :\Phi_a u_b \tilde{u}_c:
+\frac{m_b^2-m_a^2}{m_c} :A_{\mu a} A^{\mu}_b \Phi_c: \Bigr)
\, , \label{massivecoupling}
\end{equation}
for the case where all masses are nonvanishing.
But this is not the full story, because gauge invariance
can be shown to break down again at second order under certain
circumstances. As a consequence, a complete analysis of the
situation is necessary also at higher orders. In order to illustrate
the implications of gauge invariance, we present here the result
of such an analysis for the case where we have three different
colors, and we assume that at least one of them is massive
(because otherwise we would be dealing with a massless SU(2)
Yang-Mills theory).

It is found that there exist two {\em{minimal scenarios}},
which are gauge invariant also at second order,
whereas the coupling $f_{abc}$
is equal to the completely antisymmetric tensor $\varepsilon_{abc}$
for both cases:

One possibility is given by the case where two gauge boson are
equally massive, and the third one is massless. The correct
coupling structure can be obtained from (\ref{massivecoupling})
by performing a limit $m_3 \rightarrow 0$ for $m_1=m_2$.
The corresponding massless
Higgs ghost does not appear anymore in the gauge charge $Q$,
and becomes a massless physical field.
The second possibility is given by the case where all three
gauge bosons are massive, and all masses are then equal.
But it turns also out that it is necessary to introduce additional
fields in order to save gauge invariance. Indeed, it is already
sufficient to add one physical scalar to the theory (i.e. this
is the {\em{minimal}} solution): A Higgs boson.
Gauge invariance fixes the structure of the
coupling completely, but the mass of the Higgs remains
a free parameter in the theory.
The two scenarios correspond to the classical picture of
the two possible
types of spontaneous symmetry breaking for a SU(2) Yang-Mills theory:
In the first case, one couples a real SO(3) field triplet to the
gauge fields. Two degrees of freedom of this triplet
are 'eaten up' by the gauge bosons, which become massive.
In the second case, one couples a complex doublet (equivalent
to four real fields) to the
gauge bosons. Three degrees of freedom get absorbed by the gauge fields,
which become massive, and one degree of freedom remains as
a physical Higgs field. These two mechanisms are well described in many
textbooks (see e.g. \cite{Cheng}, chapter 8.3).
A detailed discussion in the present approach
of theories involving massive gauge
fields like the $Z$- and $W^{\pm}$-bosons can be found in
\cite{Mass1,Mass2,Mass3,Mass4,Mass5}.
The interesting point of the free field approach is given
by the fact that it reverses the chain of arguments usually
presented in the literature. Fields do not become massive due
to the coupling to a Higgs field, but the existence of a Higgs field
becomes necessary due to the mass of the gauge bosons
and as a consequence of perturbative gauge invariance.

\section{Quantum gravity (as massless spin 2 gauge theory)}
For the following discussion, we should always
keep in mind that quantum gravity is a nonrenormalizable
theory, and not really understood to all orders in perturbation
theory. But at least
it is possible to discuss gauge invariance of the lowest orders \cite{Schorn,Wellmann}.

The general theory of relativity can be derived from the
Einstein-Hilbert Lagrangian 
\begin{equation}
{\cal{L}}_{{\scriptscriptstyle EH}}=-\frac{2}{\kappa^2}\sqrt{-g}R
\label{eihila}
\end{equation}
where $R=g^{\mu\nu}R_{\mu\nu}$ is the Ricci scalar,
$g$ is the determinant of the metric tensor $g^{\mu\nu}$
and $\kappa^2=32\pi G$, where $G$ is Newton's gravitational
constant. It is convenient to introduce
Goldberg variables \cite{Goldberg}
\begin{equation}
\tilde{g}^{\mu\nu}=\sqrt{-g}g^{\mu\nu}.
\end{equation}
Since we are working on a perturbative level,
we expand the inverse metric $\tilde{g}^{\mu\nu}$
and assume that we have an asymptotically flat geometry
\begin{equation}
\tilde{g}^{\mu\nu}=\eta^{\mu\nu}+\kappa h^{\mu\nu}.
\end{equation}
Here $\eta^{\mu\nu}=\text{diag}(+1,-1,-1,-1)$ is the metric of
Minkowski spacetime. We used the more common symbol $g^{\mu \nu}$ for the
flat space metric in the previous sections, but use $\eta^{\mu \nu}$
in the following in order to point out the difference between $\eta$ and
$g$ more clearly.
All tensor indices will be raised and
lowered with $\eta^{\mu\nu}$. The quantity $h^{\mu\nu}$ is a
symmetric second rank tensor field, which describes gravitons
after quantization. For notational convenience, we do not care
if the indices are up or down, since the summation convention
always makes clear what is meant.
Formally (\ref{eihila}) can be expanded as an
infinite power series in $\kappa$
\begin{equation}
{\cal{L}}_{{\scriptscriptstyle EH}}=\sum_{j=0}^{\infty}\kappa^j {\cal{L}}_{{\scriptscriptstyle EH}}^{(j)}. \label{gravexpansion}
\end{equation}
The lowest order term
${\cal{L}}_{{\scriptscriptstyle EH}}^{(0)}$ is quadratic
in $h^{\mu\nu}(x)$ and defines the wave equation of
the free asymptotic fields.
The linearized Euler-Lagrange equations of motion
for $h^{\mu\nu}(x)$ are
\begin{equation}
\Box h^{\mu\nu}(x)-\frac{1}{2}\eta^{\mu\nu}\Box h(x)-h^{\mu\rho,\nu}_{{, \rho}}(x)-h^{\nu\rho,\mu}_{{, \rho}}(x)=0
\end{equation}
where $h(x)=h^{\mu}_{\mu}(x)$, and the commas denote partial derivatives
with respect to the corresponding index.
This equation is invariant
under a classical gauge transformations of the form
\begin{equation}
h^{\mu\nu}\longrightarrow h^{\mu\nu}+u^{\mu,\nu}+u^{\nu,\mu}-\eta^{\mu\nu}u^{\rho}_{{, \rho}}
\label{classicalgauge}
\end{equation}
where $u^{\mu}$ is a vector field which satisfies the wave equation
\begin{equation}
\Box u^{\mu}(x)=0.
\end{equation}
As a consequence, quantum gauge invariance for gravity will
be formulated by the help of (fermionic) vector ghosts.
This gauge transformation corresponds to the general covariance
in its linearized form of the metric tensor $g_{\mu\nu}(x)$.
The corresponding gauge condition,
compatible with (\ref{classicalgauge}) is the Hilbert-gauge,
which obviously plays a similar role as the Lorentz gauge condition for
spin 1 fields
\begin{equation}
h^{\mu\nu}_{{, \mu}}=0.
\end{equation}
Then the dynamical equation for the graviton field $h^{\mu\nu}$
reduces to the wave equation
\begin{equation}
\Box h^{\mu\nu}(x)=0.
\end{equation}
The first order term with respect to $\kappa$,
${\cal{L}}_{{\scriptscriptstyle EH}}^{(1)}$ gives
the trilinear self-coupling of the gravitons
\begin{equation}
{\cal{L}}_{{\scriptscriptstyle EH}}^{(1)}=\frac{\kappa}{2}\,h^{\rho\sigma}\Bigl(h^{\alpha\beta}_{{, \rho}}
h^{\alpha\beta}_{{, \sigma}}-\frac{1}{2}\,h^{}_{{, \rho}} h^{}_{{, \sigma}}+2\,h^{\alpha\rho}_{{, \beta}}h^{\beta\sigma}_{{, \alpha}}+
h^{}_{{, \alpha}}h^{\rho\sigma}_{{, \alpha}}
-2\,h^{\alpha\rho}_{{, \beta}}h^{\alpha\sigma}_{{, \beta}}\Bigr).
\label{gravitoncoupling}
\end{equation}

There exists many alternative derivations of this result
(\ref{gravitoncoupling}),
starting from massless tensor fields and requiring consistency
with gauge invariance in some sense. A similar point of view is the one
of Ogievetsky and Polubarinov \cite{Ogi}. They
analyzed spin 2 theories by working with a generalized
Hilbert-gauge condition to exclude the spin one part from the
outset. They imposed an invariance under infinitesimal gauge
transformations of the form
\begin{equation}
\delta h^{\mu\nu}=\partial^{\mu}u^{\nu}+\partial^{\nu}u^{\mu}+
\eta^{\mu\nu}\partial_{\alpha}u^{\alpha}
\end{equation}
and obtained Einstein's theory at the end.
Instead Wyss \cite{Wyss} considered the coupling
to matter. Then the self-coupling of the tensor-field
(\ref{gravitoncoupling}) is necessary for consistency. Wald \cite{Wald}
derived a divergence identity which is equivalent to an
infinitesimal gauge invariance of the theory. Einstein's
theory is the only non-trivial solution of this identity.
In quantum theory the problem was studied by Boulware and Deser
\cite{Deser}. These authors require Ward identities
associated with the graviton propagator to implement gauge
invariance. All authors get Einstein's theory as the unique
classical limit if the theory is purely spin two (without a
spin one admixture).

The free asymptotic field $h^{\mu \nu}$ is a symmetric tensor field
of rank two and
$u^{\mu}$ and $\tilde{u}^{\nu}$ are ghost and anti-ghost fields
on the background of Minkowski spacetime in the following.
A symmetric tensor field has ten degrees of freedom, which are more than
the five independent components of a massive spin 2 field or the
two degrees of freedom of the massless graviton field.
The additional degrees of freedom can be eliminated on the classical
level by imposing further conditions, but we will now quantize
the graviton field and introduce
the gauge charge operator $Q$. The physical states in the full
Fock space which contains eight unphysical polarizations
of the graviton and ghosts can then be characterized
directly by the help of the gauge charge as in the case of
a spin 1 field (see Appendix C, and \cite{Grillo, Grillo2,Grillo3}).

The tensor field $h^{\mu\nu}(x)$ can be quantized as a massless field
as follows
\begin{equation}
\bigl[h^{\alpha \beta}(x),h^{\mu\nu}(y)\bigr]=-ib^{\alpha
\beta \mu \nu}D_0(x-y)
\label{gravquantization}
\end{equation}
where $D_0(x-y)$ is again
the massless Pauli-Jordan distribution and the tensor
$b^{\alpha \beta \mu \nu}$ is constructed from the Minkowski metric
$\eta^{\mu\nu}$ via
\begin{equation}
b^{\alpha \beta \mu\nu}=\frac{1}{2}\bigl(\eta^{\alpha \mu}\eta^{\beta\nu}+
\eta^{\alpha \nu}\eta^{\beta \mu}-
\eta^{\alpha \beta}\eta^{\mu\nu}\bigr).
\end{equation}
An explicit representation of the field $h^{\mu\nu}(x)$
on the Fock space is given by
\begin{equation}
h^{\alpha \beta}(x)=(2\pi)^{-3/2}\int \frac{d^3 k}{\sqrt{2\omega}}
\Bigl(a^{\alpha \beta}(\vec{k}) e^{-ikx}+a^{\alpha \beta}
(\vec{k})^{\dagger} e^{+ikx}\Bigr).
\end{equation}
Here we have $\omega=|\vec{k}|$ as usual
and $a^{\alpha \beta}(\vec{k})$,
$a^{\alpha \beta}(\vec{k})^{\dagger}$ are annihilation and creation
operators on a bosonic Fock-space.
One finds that they have the following commutation relations
\begin{equation} \bigl[a^{\alpha \beta }(\vec{k}),a^{\mu\nu}(\vec{k}^{\prime})^{\dagger}
\bigr]=b^{\alpha \beta \mu\nu}\delta^{(3)}(\vec{k}-\vec{k}^{\prime}),
\end{equation}
and all other commutators vanish.

In analogy to the spin 1 theories like QCD we may introduce
a gauge charge operator by
\begin{equation}
Q:=\int\limits_{x^0=const.} d^3x \, h^{\alpha \beta }(x)_{, \beta}\overset{\leftrightarrow}{\partial_0}u_{\alpha}(x).
\label{gravgaugecharge}
\end{equation}
For the construction of the physical subspace and in order to
prove the unitarity of the $S$-matrix we must have a nilpotent
operator $Q$. Therefore we have to quantize the ghost fields
with anticommutators according to
\begin{equation}
\bigl\{u^{\mu}(x),\tilde{u}^{\nu}(y)\bigr\}=i\eta^{\mu\nu}D_0(x-y).
\end{equation}
All asymptotic fields fulfill the wave equation
\begin{equation}
\Box\, h^{\mu\nu}(x)=
\Box\, u^{\alpha}(x)=
\Box\, \tilde{u}^{\beta}(x)=0.
\end{equation}
We obtain the following commutators of the
fundamental fields:
\begin{equation}
[Q,h^{\mu\nu}] =-\frac{i}{2}\bigl(u^{\mu}_{{, \nu}}+u^{\nu}_{{, \mu}}-
\eta^{\mu\nu}u^{\alpha}_{{, \alpha}}\bigr), \label{gravcomm1}
\end{equation}
\begin{equation}
[Q,h]=iu^{\mu}_{{, \mu}},
\end{equation}
\begin{equation}
\{Q,\tilde{u}^{\mu} \}=ih^{\mu\nu}_{{, \nu}} ,
\end{equation}
\begin{equation}
\{Q,u^{\mu} \}=0.
\end{equation}
From (\ref{gravcomm1}) we immediately get
\begin{equation}
[Q,h^{\mu\nu}_{{, \mu}}]=0.
\end{equation}
The result (\ref{gravcomm1}) agrees with the
infinitesimal gauge transformations of the Goldberg variables,
so that our quantization and choice
of $Q$ corresponds to the classical framework.

Again, first order gauge invariance means that $[Q,T_1]$ is a divergence
in the sense of vector analysis, i.e.
\begin{equation}
[Q,T_1(x)]=i\partial_{\mu}T_{1/1}^{\mu}(x),
\end{equation}
and definition of the $n$-th order gauge invariance then reads
\begin{equation}
[Q,T_n]=\bigl[Q,T_n\bigr]=i\sum_{l=1}^{n}\frac{\partial}{\partial x^{\mu}_l}T_{n/l}^{\mu}(x_1,\ldots,x_l,\ldots,x_n).
\end{equation}
Like in the case of QCD, we are forced to add a ghost part
to the self-coupling term of the gravitons.
The first order graviton coupling according to (\ref{gravitoncoupling})
is
\begin{equation}
\tilde{T}_1^{h}=\frac{i \kappa}{2}\,h^{\rho\sigma}
\Bigl(h^{\alpha\beta}_{{, \rho}}
h^{\alpha\beta}_{{, \sigma}}-\frac{1}{2}\,h^{}_{{, \rho}} h^{}_{{, \sigma}}+2\,h^{\alpha\rho}_{{, \beta}}h^{\beta\sigma}_{{, \alpha}}+
h^{}_{{, \alpha}}h^{\rho\sigma}_{{, \alpha}}
-2\,h^{\alpha\rho}_{{, \beta}}h^{\alpha\sigma}_{{, \beta}}\Bigr).
\label{gc1}
\end{equation}
By adding a physically irrelevant divergence to
$\tilde{T}_1^{h}$, we obtain
a more compact expression for the graviton interaction
\begin{equation}
T_1^h(x)=i \kappa :\Bigl( \frac{1}{2} h^{\mu \nu}
h^{\alpha \beta}_{, \mu} h^{\mu \nu}
h^{\alpha \beta}_{, \nu}+ h^{\mu \nu} h^{\nu \alpha}_{, \beta}
h^{\mu \beta}_{, \alpha}-\frac{1}{4} h^{\mu \nu} h_{, \nu} h_{, \nu}
\Bigr):.
\end{equation}
The ghost coupling turns out to be the one first suggested
by Kugo and Ojima \cite{Kugograv}, namely
\begin{equation}
T_1^u=i \kappa : \tilde{u}^\nu_{, \mu} \Bigl( h^{\mu \nu}_{, \rho}
u^\rho - h^{\nu \rho} u^{\mu}_{, \rho} - h^{\mu \rho}
u^\nu_{, \rho} + h^{\mu \nu} u^\rho_{, \rho} \Bigr) : . \label{gc2}
\end{equation}
It is a nice detail that the four graviton vertex 
which follows from
the second order term in (\ref{gravexpansion})
is also proliferated by gauge invariance at second order.
It is therefore quite probable that also all higher vertex couplings
which appear in (\ref{gravexpansion}) are proliferated by quantum gauge
invariance. It is also possible to derive the couplings
(\ref{gc1}),(\ref{gc2}) from perturbative gauge invariance
similarly as it was done in sect. 6 for Yang-Mills theories.
For this and further reading we refer to
the recent monograph \cite{Ghoststory}.

We finally point out that it would be interesting to analyze
the Higgs mechanism for massive gravity \cite{Chamse}
in the framework presented above, also in connection with
the dark matter problem. A first step in this direction is
presented in \cite{MassiveGravity}.

\section{Appendices}
\subsection*{Appendix A: Commutator of the gauge charge $Q$
with gauge fields}
First we derive the distributional identity (\ref{timederiv}).
The Pauli-Jordan distribution has an integral representation
\begin{equation}
D(x)=\frac{i}{(2 \pi)^3} \int d^4 k \,
\delta(k^2) \mbox{sgn}(k_0) e^{-i kx}
\end{equation}
Using the identity
\begin{equation}
\delta(k^2)=\delta(k_0^2-\vec{k}^{\, 2})=\frac{1}{2 |k^0|}
\Bigl( \delta(k^0-|\vec{k}|) + \delta(k^0+|\vec{k}|) \Bigr),
\end{equation}
we obtain (mind the $\mbox{sgn}(k^0)$)
\begin{displaymath}
\partial_0 D(x)= \frac{i}{(2 \pi)^3} \int
\frac{d^4 k}{2 |k^0|}
\Bigl( \delta(k^0-|\vec{k}|) - \delta(k^0+|\vec{k}| \Bigr)
(-i k^0) e^{-ikx}
\end{displaymath}
\begin{equation}
\frac{1}{2 (2 \pi)^{3}} \int d^3 k \,
\Bigl( e^{-i(|\vec{k}|x^0 -\vec{k} \vec{x})} +
e^{-i(-|\vec{k}|x^0 -\vec{k} \vec{x})} \Bigr).
\end{equation}
Restricting this result to $x^0=0$, we get
the desired result (\ref{timederiv})
\begin{equation}
\partial_0 D(x) |_{x^0=0} =(2 \pi)^{-3} \int d^3 k \,
e^{+i\vec{k} \vec{x}} = \delta^{(3)} (\vec{x}).
\end{equation}
In a completely analogous way, one derives
\begin{equation}
\partial_0^2 D(x) |_{x^0=0} = 0, \quad {\vec{\nabla}}
D(x) |_{x^0=0}=0. \label{deriv2}
\end{equation}
Note that we always consider the well-defined differentiated
distribution first, which gets then restricted to a subset
of its support.

As an example, we investigate now the commutator
$[Q, A_\mu(y)]$, and we omit the trivial color index.
The commutator is given explicitly by
\begin{equation}
[Q, A_\mu(y)]= [\int \limits_{x^0=y^0} d^3 x \,
\partial_\nu A^\nu (x) \stackrel
{\leftrightarrow}{\partial_0^x} u(x), A_\mu(y)]=
i\int \limits_{x^0=y^0} d^3 x \, \partial_\mu^x D(x-y)
\stackrel {\leftrightarrow}{\partial_0^x} u(x).
\end{equation}
Here, we made use of the freedom to choose any constant value
for $x^0$, and therefore we set $x^0=y^0$, such that we
have $x^0-y^0=0$ and we can apply (\ref{timederiv}, \ref{deriv2})
in the sequel.

For $\mu =0$, we have
\begin{displaymath}
[Q, A_0(y)]= 
i\int \limits_{x^0=y^0} d^3 x \, \partial_0^x D(x-y)
\stackrel {\leftrightarrow}{\partial_0^x} u(x)
\end{displaymath}
\begin{equation}
=i \int \limits_{x^0=y^0} d^3 x \, \delta^{(3)} (\vec{x}-\vec{y})
\partial_0^x u(x) = i \partial_0 u(y),
\end{equation}
where we have used that the double timelike derivative of
$D$ vanishes on the integration domain according to (\ref{deriv2}).
The result for the commutator of $Q$ with the spacelike
components of $A$ is also obtained by using
(\ref{timederiv}, \ref{deriv2})
and by shifting the gradient acting of the Pauli-Jordan distribution
by partial integration on the ghost field.

\subsection*{Appendix B: The Becchi-Rouet-Stora-Tyutin transformation
and its free field version}
The gluon vector potential can be represented by the traceless hermitian
$3 \times3$ standard Gell-Mann matrices $\lambda^a$, $a=1,...8$
\begin{equation}
A_\mu=\sum_{a=1}^{8} A_\mu^a \frac{\lambda^a}{2} =: A_\mu^a
\frac{\lambda^a}{2} .
\end{equation}
The $\lambda$'s satisfy the commutation and normalization relations
\begin{equation}
\Bigl[ \frac{\lambda^a}{2}, \frac{\lambda^b}{2} \Bigr]=
i f_{abc} \frac{\lambda^c}{2}, \quad \mbox{tr} \, (\lambda^a \lambda^b)=
2 \delta_{ab}, \label{goodrelations}
\end{equation}
and the numerical values of the structure constants
$f_{abc}=-f_{bac}=-f_{acb}$ can be found
in numerous QCD textbooks.
Since we are working with a fixed matrix representation, we do not care
whether the color indices are upper or lower indices.

The natural generalization of the QED Lagrangian to the
Lagrangian of purely gluonic QCD is
\begin{equation}
{\cal{L}}_{gluon} = -\frac{1}{2} \mbox{tr} \, G_{\mu \nu} G^{\mu \nu}
=-\frac{1}{4} G^a_{\mu \nu} G_a^{\mu \nu},
\end{equation}
with
\begin{equation}
G_{\mu \nu}=\partial_\mu A_\nu-\partial_\nu A_\mu-ig[A_\mu,A_\nu]
\end{equation}
or, using the first relation of (\ref{goodrelations})
\begin{equation}
G_{\mu \nu}^a=\partial_\mu A_\nu^a - \partial_\nu A_\mu^a+
g f_{abc} A_\mu^b A_\nu^c .
\end{equation}
It is an important detail that we are working with {\em{interacting}}
classical fields here, therefore the field strength tensor contains
a term proportional to the coupling constant in contrast to the free
field tensor
$F_{\mu \nu}^{free}=\partial_\mu A_\nu^{free}-\partial_\nu A_\mu^{free}$
used in this paper. ${\cal{L}}_{gluon}$ is invariant under
classical local gauge transformations
\begin{equation}
A_\mu(x) \rightarrow U(x) A_\mu(x) U^{-1}(x)
+\frac{i}{g} U(x) \partial_\mu U^{-1}(x),
\end{equation}
where $U(x) \in SU(3)$.

We extract now the first order gluon coupling from the Lagrangian.
The Lagrangian
\begin{equation}
{\cal{L}}_{gluon}=-\frac{1}{4} [\partial_\mu A_\nu^a - \partial_\nu A_\mu^a+
g f_{abc} A_\mu^b A_\nu^c]
[\partial^{\mu} A^{\nu}_{a} - \partial^{\nu} A^{\mu}_{a}+
g f_{ab'c'} A^{\mu}_{b'} A^{\nu}_{c'}],
\end{equation}
contains obviously the free field part (this terminology is not
really correct, since we are dealing with interacting fields
here)
\begin{equation}
{\cal{L}}_{gluon}^{free}=-\frac{1}{4} [\partial_\mu A_\nu^a - \partial_\nu A_\mu^a][\partial^{\mu} A^{\nu}_{a} - \partial^{\nu} A^{\mu}_{a}]
\end{equation}
and the first order interaction part is given by
\begin{displaymath}
{\cal{L}}_{gluon}^{int}=
-\frac{1}{4} [\partial_\mu A_\nu^a - \partial_\nu A_\mu^a]
[g f_{ab'c'} A^{\mu}_{b'} A^{\nu}_{c'}]-\frac{1}{4}
[g f_{abc} A_\mu^b A_\nu^c]
[\partial^{\mu} A^{\nu}_{a} - \partial^{\nu} A^{\mu}_{a}]
\end{displaymath}
\begin{displaymath}
=-\frac{g}{2} f_{abc} A_\mu^b A_\nu^c
[\partial^{\mu} A^{\nu}_{a} - \partial^{\nu} A^{\mu}_{a}]=
-\frac{g}{2} f_{abc} A_\mu^a A_\nu^b
[\partial^{\mu} A^{\nu}_{c} - \partial^{\nu} A^{\mu}_{c}]
\end{displaymath}
\begin{equation}
=g f_{abc} A_\mu^a A_\nu^b \partial^{\nu} A^{\mu}_{c}.
\end{equation}
From this term follows the first order interaction ansatz
$T_1^A=i :{\cal{L}}_{gluon}^{int}:$ (\ref{gluonQCD}).

Since we are working in Feynman gauge, we add
the corresponding gauge fixing term ${\cal{L}}_{gf}$
to the Lagrangian. Additionally,
we add a ghost term which leads to the ghost interaction
(\ref{ghostQCD}). The total Lagrangian then reads
\begin{displaymath}
{\cal{L}}_{QCD}={\cal{L}}_{gluon}+{\cal{L}}_{gf}
+{\cal{L}}_{ghost}
\end{displaymath}
\begin{equation}
={\cal{L}}_{gluon}-\frac{1}{2}(\partial_\mu A^\mu_a)^2+
\partial^\mu \tilde{u} ( \partial_\mu u_a-g f_{abc} u_b A_{\mu c}).
\end{equation}
The classical ghosts are anticommuting Grassmann numbers, i.e.
$u^2=\tilde{u}^2=0, u \tilde{u}=-\tilde{u} u$.

The BRST transformation is defined by
\begin{equation}
\delta A_\mu^a= i\lambda (\partial_\mu u_a-g f_{abc} u_b A_{\mu c}),
\end{equation}
\begin{equation}
\delta \tilde{u}_a=-i \lambda \partial_\mu A^\mu_a,
\end{equation}
\begin{equation}
\delta u_a= \frac{g}{2} \lambda f_{abc} u_b u_c,
\end{equation}
where $\lambda$ is a space-time independent anticommuting
Grassmann variable.
The special property of the BRST transformation is the fact that
the actions
\begin{equation}
S_{gluon}=\int d^4 x \, {\cal{L}}_{gluon}, \quad
S_{gf}+S_{ghost}=\int d^4 x \, ({\cal{L}}_{gf}
+{\cal{L}}_{ghost})
\end{equation}
and $S_{total}=S_{gluon}+S_{gf}+S_{ghost}$
are all invariant under the transformation:
\begin{equation}
\delta S_{gluon}=0, \quad \delta(S_{gf}+S_{ghost})=0.
\end{equation}

The similarity of free quantum gauge transformation introduced
in this paper to the BRST transformation is obvious.
One important difference is the absence of interaction terms
$\sim g$. Furthermore, the free quantum gauge transformation
is a transformation of free quantum fields, whereas the
BRST transformation is a transformation of classical fields,
which enter in path integrals when the theory is
quantized. Finally, the free gauge transformation leaves
the $T_n$'s invariant up to divergences, whereas the
BRST transformation is a symmetry of the full QCD Lagrangian.
How the two symmetries are intertwined perturbatively is explained
in \cite{Hurth1}. A more rigorous axiomatic approach is discussed in
\cite{MasterWard,Dutsch1}.

\subsection*{Appendix C: Some technical remarks concerning
the gauge charge Q}
Using the Leibnitz rule for graded algebras gives for the
gauge charge for massless spin 1 fields
\begin{displaymath}
Q^2=\frac{1}{2} \{Q,Q\} =\frac{1}{2} \int \limits_{x_0=const.}
d^3 x \partial_\nu A^\nu_a(x) \{  \stackrel
{\leftrightarrow}{\partial_{x_0}} u_a(x) , Q \}  \label{nr22}
\end{displaymath}
\begin{equation}
-\frac{1}{2}  \int \limits_{x_0=const.}
d^3 x [\partial_\nu A^\nu_a(x) , Q] \stackrel{\leftrightarrow}{\partial_{x_0}}
u_a(x) = 0 \quad ,
\end{equation}
i.e.~Q is nilpotent. This basic property of Q, which holds
also in the massive case, and the
so-called Krein structure on the
Fock-Hilbert space \cite{14,15} allows to prove unitarity of the
$S$-matrix on the physical Hilbert space ${\cal{F}}_{phys}$, which is a subspace of the
Fock-Hilbert space ${\cal{F}}$ containing also the unphysical ghosts and
unphysical degrees of freedom of the vector field \cite{6}.

The physical Fock space can be expressed by the kernel and the
range of Q \cite{6,14}, namely
\begin{equation}
F_{phys}=ker \, Q / ran \, Q=ker\{Q,Q^\dagger\} \quad . \label{physspace}
\end{equation}
Again, this characterization of the physical space holds also for the
massive spin 1 case, and for the gauge charge (\ref{gravgaugecharge})
used for gravity.
The Krein structure is defined by introducing a conjugation $K$
\begin{equation}
a_0(\vec{k})^K = -a_0(\vec{k})^\dagger ,
\quad a_j(\vec{k})^K=a_j(\vec{k})^\dagger ,
\end{equation}
so that $A_\mu^K=A_\mu$, and on the ghost sector
\begin{equation}
c_2(\vec k)^K=c_1(\vec k)^\dagger,\quad c_1(\vec k)^K=
c_2(\vec k)^\dagger ,
\end{equation}
so that $u^K=u$ is $K$-selfadjoint and $\tilde u^K=-\tilde u$. Then
$Q$ is densely defined on the Fock-Hilbert space
and becomes $K$-symmetric $Q
\subset Q^K$. Roughly speaking, the K-conjugation is the
natural generalization of the usual hermitian conjugation
to the full (unphysical) Fock space.

A calculation shows that
the anticommutator in (\ref{physspace}) is essentially the number
operator for unphysical particles
\begin{equation}
\{Q^\dagger,Q\} = 2 \int d^3 k \, \vec{k}^2
\bigl[ b_1^\dagger(\vec{k}) b_1(\vec{k}) +
b_2^\dagger(\vec{k})b_2(\vec{k})+c_1^\dagger(\vec{k})c_1(\vec{k})+
c_2^\dagger(\vec{k})
c_2(\vec{k}) \bigr] ,
\end{equation}
with
\begin{equation}
b_{1,2}=(a_{\|} \pm a_0)/\sqrt{2} \quad , \quad
a_{\|}=k_ja_j/|\vec{k}| ,
\end{equation}
which implies (\ref{physspace}).

The nilpotency of $Q$ allows for standard homological notions \cite{6}:
Consider the field algebra $\cal{F}$ consisting of the polynomials in the
(smeared) gauge and ghost fields and their Wick powers. 
Defining a gauge variation for a Wick monomial $F$ according to
\begin{equation}
d_Q F \stackrel{def}{=} QF - (-1)^{n_F} FQ  ,
\end{equation}
where $n_F$ is the number of ghost fields in F,
Q becomes a differential operator in the sense of homological algebra,
and we have
\begin{equation}
d_Q^2=0 \Longleftrightarrow \{Q,[Q,F_b]\} = [Q,\{Q,F_f\}]=0  ,
\end{equation}
where $F_b$ is a bosonic and $F_f$ a fermionic operator
and $d_Q(FG)=(d_QF)G+(-1)^{n_F}Fd_QG$.
For example, we get
\begin{equation}
d_Q :A_{\mu a} u_b \partial^\mu \tilde{u}_c: = :[Q,A_{\mu a}]u_b
\partial^\mu \tilde{u}_c: + :A_{\mu a}[Q,u_b] \partial^\mu \tilde{u}_c:
-:A_{\mu a}u_b \{Q, \partial^\mu \tilde{u}_c\}: \quad .
\end{equation}
If $F=d_QG$, then $F$ is called a coboundary.
The term (\ref{cobo})
\begin{equation}
\tilde f_{abc}:\partial_\nu A^\nu_a u_b {\tilde{u}}_c:=\frac{i}{2} d_Q
(\tilde f_{abc}: {\tilde{u}}_a u_b {\tilde{u}}_c:) \, .
\end{equation}
is such a coboundary, which does not
contribute to the physical QCD S-matrix.

\section*{Appendix D: Gauge invariance at n-th order}
A short discussion is given here which explains how the
simple condition of gauge invariance at n-th order
in the coupling constant (\ref{dive}) emerges
as a generalization from the first order condition (\ref{gi}).

A thorough mathematical treatment of the subject would involve a
discussion of operator valued distributions, which is not given
here for the sake of brevity. Nevertheless, all mathematical steps
presented in the following can be put on a sound mathematical basis.
It is useful to remind some basic facts:
Free quantum fields like the free scalar field $\varphi(x)$
(\ref{freefield}) are operator valued distributions on a
Fock-Hilbert space, i.e. an operator $\varphi(g)$ is
obtained after smearing out the field operator $\varphi(x)$
with a test function $g(x)$.
This can be written {\em{formally}} as
\begin{equation}
\varphi(g) =\int d^4 x \, \varphi(x) g(x),
\end{equation}
where $g \in {\cal{S}}({\bf{R}}^4)$ is in the Schwartz space
of infinitely differentiable and rapidly decreasing
test functions.
Furthermore, the tensor product $\varphi(x) \psi(y)$
of free fields (and normally ordered products of free fields)
is also an operator valued distribution,
whereas the local product $\varphi(x)\varphi(x)$ is only defined
after normal ordering. The 'value' of a distribution in
a single point makes no sense in general. It is
sometimes convenient to treat distributions like ordinary
functions, as we will do it below, because in many cases
the simplified insight gained from such a simplification
allows to construct a full mathematical proof at a later stage.

As mentioned in the paper, the $T_n(x_1,...x_n)$ are
well-defined time-ordered products of the first
order coupling $T_1(x)$, and they are expressed in terms of Wick
monomials of free fields. The construction of the $T_n$ requires some
care:
If the arguments $(x_1,...x_n)$ are all time-ordered, i.e.
if we have
\begin{equation}
x_1^0>x_2^0>...>x_n^0 ,
\end{equation}
then $T_n$ is simply given by
\begin{equation}
T_n(x_1,...x_n) = T_1(x_1)T_1(x_2) ... T_1(x_n)
\end{equation} 
According to the definition (\ref{smatrix}), the
$T_n(x_1,...x_n)$ can be considered symmetric in $x_1,...x_n$.
Using this fact
allows us in principle to obtain the operator-valued distribution
$T_n$ inductively
everywhere except for the 'complete diagonal' $\Delta_n=\{
x_1=...=x_n\}$ \cite{Stora}. The construction is inductive since
in the construction of $T_n$ subdiagrams with lower order appear.
If $T_n$ were a C-number distribution, 
we could make it a well-defined
distribution for all $x_1,...x_n$ by extending the distribution
from ${\cal{R}}^{4n}/\Delta_n$ to ${\cal{R}}^{4n}$.
In the case of free field operators,
the problem can be reduced to a C-number problem by the Wick
expansion of the operator-valued distributions.
The extension $T(x_1,...x_n)$ is, of course, not unique: it is ambiguous
up to distributions with local support $\Delta_n$. This ambiguity
can be further reduced by the help of symmetries (in particular gauge 
invariance) and power counting theory, and it is this local ambiguity
which shows up as ultraviolet divergences in Feynman diagram
calculations.

The concrete inductive construction of the $T_n$
is discussed in detail in a famous paper of Epstein and
Glaser \cite{Epstein} for scalar field theories,
and explained in a more pedagogical way in \cite{Electro}.
For many practical calculations, it is advantageous to work
in momentum space since the distributions, i.e. the Green's
functions which occur in the calculations, behave much smoother
in $p$-space than in $x$-space. The work of Epstein and Glaser
is based on a treatment in real space, whereas in \cite{Electro}
practical calculations are performed in momentum space.
An advantage of the real space formulation is also that it allows
to formulate a consistent renormalization theory on a curved
physical background \cite{Brunetti}. Interesting topics like
scale invariance, renormalization and the renormalization group
can also be treated in real space \cite{Zhang,Prangediff,Grigorescale}.

As a specific example, we consider the second order contribution
to the S-matrix which is given by
\begin{equation}
S_2=\frac{1}{2 !}
\int d^4 x \, d^4 y \, T[T_1(x) T_1(y)]
=\frac{1}{2}
\int d^4 x  \, d^4 y \, T_2(x,y). \label{s2matrix}
\end{equation}
We consider first the simple product $T_1(x) T_1(y)$
without time ordering.
Then we have 
\begin{displaymath}
[Q,T_1(x) T_1(y)]=[Q,T_1(x)] T_1(y)+T_1(x) [Q,T_1(y)]
\end{displaymath}
\begin{equation}
= i \partial_\mu T^\mu_{1/1}(x) T_1(y)+
i T_1(x) \partial_\mu T^\mu_{1/1}(y).
\end{equation}
Considering the time-ordered product $T[T_1(x) T_1(y)]$,
we have to distinguish three cases. In the first case, we
have $x^0 > y^0$, and therefore $T[T_1(x) T_1(y)]=T_1(x)T_1(y)$.
The second, a bit less trivial case is given for
$x^0 < y^0$. Then we have $T[T_1(x) T_1(y)]=T_1(y)T_1(x)$, and
\begin{displaymath}
[Q,T(T_1(x) T_1(y))]=[Q,T_1(y)] T_1(x)+T_1(y) [Q,T_1(x)]
\end{displaymath}
\begin{displaymath}
= i \partial_\mu T^\mu_{1/1}(y) T_1(x)+
i T_1(y) \partial_\mu T^\mu_{1/1}(x)
\end{displaymath}
\begin{equation}
=i T[\partial_\mu T^\mu_{1/1}(x) T_1(y)]+
i T[T_1(x) \partial_\mu T^\mu_{1/1}(y)].
\end{equation}
Finally, we may have $x^0 = y^0$. If $(\vec{x}-\vec{y}) \neq \vec{0}$,
i.e. if $x$ and $y$ are spacelike separated, then we may perform
'a little' Lorentz transformation such that $x^0 > y^0$, and
the discussion above applies.
If the special case that $x=y$, gauge invariance is not trivially
fulfilled. This is not astonishing, since this is the case
on the diagonal $\Delta_2=\{x=y\}$ where
perturbation theory may fail to calculate time-ordered products
of the first order interaction. Requiring that gauge invariance
holds also for $x=y$ leads to Ward-Takahashi identities in QED,
and to the so-called Slavnov-Taylor identities for QCD.

The simplest example for such an identity is obtained from the
vacuum polarization contribution to the S marix
at second order, which can be written
in the form
\begin{equation}
T_2^{vp}(x,y) = :A_\mu(x) t^{\mu \nu}(x-y) A_\nu(y):,
\end{equation}
where $t^{\mu \nu}(x-y)=t^{\nu \mu}(x-y)$ is a C-number distribution.
Operator gauge invariance requires
\begin{displaymath}
[Q,T_2^{vp}(x,y)]=i \partial_\mu u(x) t^{\mu \nu}(x-y) A_\nu(y)
+i A_\mu (x) t^{\mu \nu}(x-y) \partial_\nu u(y)
\end{displaymath}
\begin{equation}
=i \partial_\mu^x  [u(x) t^{\mu \nu}(x-y) A_\nu(y)]
+i \partial_\nu^y [A_\mu (x) t^{\mu \nu}(x-y)  u(y)],
\end{equation}
and therefore
\begin{equation}
\partial_\mu^x t^{\mu \nu}(x-y)=\partial_\nu^y t^{\mu \nu}(x-y)=0.
\end{equation}
In momentum space, we get the C-number identity
\begin{equation}
k_\mu {\hat{t}}^{\mu \nu}(k)=k_\nu {\hat{t}}^{\mu \nu}(k)=0,
\end{equation}
hence the vacuum polarization tensor is transversal.

\subsection*{Appendix E: The adiabatic limit}
The perturbative expression (\ref{smatrix}) is problematic,
because the time-ordered products $T_n$
are operator valued distributions after regularization,
and they have to be smeared out by test functions.
In order to be more precise in a
mathematical sense,  we introduce a test function
$g_0(x) \in {\cal{S}}({\bf{R}}^4)$ with $g_0(0)=1$ and replace
expression (\ref{smatrix}) by
\begin{equation}
S={\bf{1}}+\sum \limits_{n=1}^{\infty} \frac{1}{n !}
\int d^4 x_1 \ldots d^4 x_n \, T_n(x_1, \ldots x_n)
g_0(x_1)...g_0(x_n).
\end{equation}
Here, $g_0$ acts as an infrared regulator, which switches
off the long range part of the interaction in theories where
massless fields are involved.
E.g., in QED the emission of soft photons is switched
by $g_0$, and as long as we do not perform a so-called
adiabatic limit
$g_0 \rightarrow 1$, the matrix elements of the S-matrix
remain finite.
One possibility to perform the adiabatic limit is by
scaling the switching function $g_0(x)$, i.e. one replaces
$g_0(x)$ by $g(x)=g_0(\epsilon x)$ and performs the
limit $\epsilon \rightarrow 0$, such that $g$ approaches
everywhere the value $1$.
If the S-matrix is modified by a gauge transformation,
operators which are divergences are
added to the n-th order term $T_n$. Such a contribution
can be written as
\begin{displaymath}
\int d^4 x_1 ... d^4 x_n \, 
\partial_\mu^{x_l} O^{... \mu... }(x_1,...,x_l,...x_n) g(x_1) ...
g(x_l) ...g(x_n)
\end{displaymath}
\begin{equation}
=-\int d^4 x_1 ... d^4 x_n \, 
 O^{... \mu... }(x_1,...,x_l,...x_n) g(x_1) ...
\partial_\mu^{x_l} g(x_l) ...g(x_n). \label{farout}
\end{equation}
In the adiabatic limit, the gradient $\partial_\mu^{x_l} g(x_l)$
vanishes. Unfortunately, this property of the scaling limit
does not guarantee that the whole term (\ref{farout}) vanishes
(see also \cite{Infra1,Infra2}).

The infrared problem is not
really understood in QCD, and all proofs of unitarity
which exist in the literature have to be taken with a grain
of salt, because they are avoiding the discussion of infrared
problems somehow.

A thorough perturbative approach to the construction of
the {\it local} algebras of observables which avoids
the adiabatic limit is given for QED in \cite{MasterWard}
and (under the assumption that there are no anomalies) for
non-Abelian gauge theories in \cite{Dutsch1}.

\section*{Acknowledgements}
I thank Dirk Trautmann and Florian Weissbach for 
carefully reading the manuscript.
This paper is based on talks held at the
Institute for Theoretical Physics at the University
of Trento, Italy, and the Center for Theoretical Physics
at the CNRS in Marseille, France.

The work was supported by the Swiss National Science Foundation.

\end{document}